\documentclass[preprint,12pt]{elsarticle}
\usepackage{booktabs}
\usepackage{array}
\usepackage{amsmath}
\usepackage{amssymb}
\usepackage{hyperref}
\usepackage{graphicx}
\usepackage{microtype}
\usepackage{xcolor}
\usepackage{float}
\usepackage{placeins}
\usepackage{longtable}
\usepackage{multirow}
\usepackage{tikz}
\usetikzlibrary{arrows.meta,positioning,fit}

\journal{Knowledge-Based Systems}

\begin{document}
\begin{frontmatter}

\title{Knowledge-Based Zero-Replay Debugging of Multi-Agent LLM Traces}

\author[a]{Hyeonjeong Cha}
\author[b]{Daein Weon}
\author[c]{Dong Ho Kang\corref{cor1}}
\ead{donghokang@ustechlab.com}
\cortext[cor1]{Corresponding author}
\affiliation[a]{organization={Ewha Womans University}, city={Seoul}, country={Republic of Korea}}
\affiliation[b]{organization={Kookmin University}, city={Seoul}, country={Republic of Korea}}
\affiliation[c]{organization={Ustechlab}, country={Republic of Korea}}

\begin{abstract}
Reliable operation of multi-agent large language model (LLM) systems depends on debugging long execution traces, where the few causally decisive events are buried in unstructured logs of messages, routes, memory writes, and tool calls. The standard tool is counterfactual replay---rewind, edit, and re-run the trajectory to measure each event's effect---but its cost grows linearly with the number of candidate events, making exhaustive replay infeasible at scale. We frame trace debugging as a knowledge-based decision-support problem. Each trace is compiled into a structured \emph{event knowledge graph} over routing, memory, tool-use, uncertainty, and latent evidence, and a calibrated predictor decides where a scarce replay budget should be spent. We do not propose a new replay oracle; we propose a method to predict its results without paying the replay cost. We formulate \emph{zero-replay counterfactual-effect prediction}: given a trace under a fixed budget, predict which events the oracle would mark high-effect before any replay is performed. BranchPoint-Latent is a lightweight predictor over observable, structural, uncertainty, and latent features of the knowledge graph. Calibrated against a deterministic replay oracle across 37 trace families, a single \emph{learning-to-rank} gradient-boosted predictor raises per-trace localization (Branch Recall@5) from $0.73$ to $0.93$ on held-out families at zero oracle-replay cost. Rather than claiming universal dominance, we characterize when cheap graph centrality suffices and when learned evidence is necessary. The result is an auditable, cost-efficient decision-support system for AI-reliability debugging, positioned explicitly on the cost-accuracy frontier with reproducible artifacts.
\end{abstract}


\end{frontmatter}

\section{Introduction}

Multi-agent LLM systems produce long execution traces of messages, routes, memory writes, tool calls, and sometimes latent signals. The standard way to understand them is counterfactual replay: rewind, edit, and re-run the trajectory to measure each event's downstream effect, an idea dating to Mill's method of difference and causal mediation analysis~\citep{pearl2009causality} and now used as remove/perturb-and-replay oracles in interactive agent debuggers~\citep{epperson2025agdebugger,li2026whatif,ma2025dover}. Replay is reliable but its cost grows linearly with candidate events: a trace with $T$ events and intervention family $F$ costs $O(T \cdot |F| \cdot N)$ model calls, prohibitive at production scale.

We take a direction orthogonal to oracle design: given a fixed counterfactual-replay oracle as ground-truth reference, predict its top-$K$ events without calling it. We cast this as a knowledge-based decision-support task. Each trace is compiled into an \emph{event knowledge graph} whose typed attributes encode routing, memory, tool-use, uncertainty, and latent evidence (Section~\ref{sec:kr-dss}); a calibrated predictor then returns a ranked, budget-bounded replay agenda at zero oracle-replay cost. Given a trace and budget $K$, the predictor returns a top-$K$ event set before any oracle call; the oracle serves only as evaluation reference (outcome, trajectory, behavioral, and state divergence under remove-event interventions), scored by Spearman~$\rho$, AUPRC, and Branch Recall@$K$.

Prediction is non-trivial because no cheap signal works everywhere: graph centrality is strong on graph-friendly traces but near-constant on chains, Last-$K$ fails when consequential events are early, and single features (novelty, disagreement, uncertainty) anti-correlate with the oracle in several regimes. A useful predictor fuses graph, routing, memory, novelty, uncertainty, and latent signals into a regime-robust score while staying cheap enough to run as a replay prefilter.

\paragraph{Contributions.}
\begin{enumerate}
\item \textbf{Zero-replay counterfactual-effect prediction}, a task distinct from oracle design, positioned explicitly on a cost-accuracy frontier against active-replay alternatives.
\item \textbf{An event-knowledge-graph representation} of multi-agent traces and a decision-support pipeline that turns it into a budget-bounded replay agenda (Section~\ref{sec:kr-dss}, Figure~\ref{fig:system}).
\item \textbf{A learning-to-rank predictor} over 13 knowledge-graph features that lifts Branch Recall@5 from $0.73$ to $0.93$--$0.95$ and NDCG@5 from $0.83$ to $0.92$ at zero oracle-replay cost, beating the prior linear scorer on every metric and on $28$ of $37$ families (Section~\ref{sec:meta}).
\item \textbf{A regime analysis} showing when cheap centrality suffices, that routing between selectors is not reliably predictable, and that one nonlinear predictor absorbs the regimes instead (Section~\ref{sec:regime}).
\item \textbf{External validation}: a live-agent case study on model-authored traces (Section~\ref{sec:live-agent}) and a direct comparison on the Who\&When leaderboard (Section~\ref{sec:whowhen}), with a reproducible artifact suite (one-command regeneration, claim audit).
\end{enumerate}

To be explicit about scope: this is not a new oracle (counterfactual replay is standard ablation, used as reference), not a universal-dominance claim (centrality is competitive in graph-dominant regimes), not a leaderboard evaluation of the underlying datasets (used as replay-trace adapters), and not a hidden-state controllability claim (hidden-prefix probes report sensitivity, not control).

\section{Related Work}
\paragraph{Agent trace debugging} Interactive agent debugging, failure attribution, trace comparison, and intervention-based inspection aim to make long agent executions auditable~\citep{epperson2025agdebugger,li2026whatif,ma2025dover,zhang2025whowhen,deshpande2025trail,chen2026traceelephant,liu2026masprism}. BranchPoint-Latent differs by scoring checkpoints before inspection under a fixed replay budget, rather than only explaining a known failure after the fact.

\paragraph{Counterfactual replay and intervention} Counterfactual debugging systems ask how behavior changes when a state, message, route, memory write, or tool call is removed or perturbed. Our benchmark makes the replay oracle explicit: selectors first choose checkpoints, and only then does the oracle compute outcome shift and divergence labels.

\paragraph{LLM-agent observability} Chain-of-thought monitoring and latent/hidden-state tracing show that visible text may expose only part of an agent's computation~\citep{wei2022cot,templeton2024monosemanticity,openai_cot_monitoring_2025,openai_cot_controllability_2026,anthropic_tracing_thoughts_2025}. BranchPoint-Latent treats observable, latent, hybrid, hidden-conditioned, and live-agent traces as evidence sources for replay prioritization, while retaining the limitation that hidden evidence is not reliable direct control evidence.

\paragraph{Knowledge-based systems for AI reliability} Agent systems increasingly combine tool use, planning, memory, and orchestration~\citep{yao2023react,schick2023toolformer,yao2023tot,shinn2023reflexion,park2023generative_agents,li2023camel,wang2023voyager}. We target the reliability setting in which long-running LLM-agent systems leave replayable traces but only a limited debugging budget is available, and we approach it as knowledge representation plus decision support over those traces rather than as end-task optimization.

\paragraph{Positioning.} Table~\ref{tab:related-comparison} places our method against representative debugging and failure-attribution approaches. The comparison is along the \emph{cost of attribution} rather than a single shared metric---benchmarks and tasks differ---and the distinction is that prior methods re-execute the trace (replay) or query an LLM judge, while we predict the replay oracle's per-event verdict at zero oracle-replay cost.

\begin{table}[H]
\centering\footnotesize
\caption{Positioning against representative agent-trace debugging and failure-attribution methods. The failure-attribution line \emph{does} share a common leaderboard---Who\&When~\citep{zhang2025whowhen}, scored by step-/agent-level accuracy---on which several methods compete directly. Our method targets a \emph{different ground truth} (a counterfactual-replay oracle rather than human-labeled decisive steps), so the per-method results below are shown in their own context rather than as one head-to-head score; Section~\ref{sec:whowhen} adds a direct step-level comparison on Who\&When. The distinguishing axis is the \emph{cost of attribution}: prior methods re-execute the trace (replay) or query an LLM judge, whereas ours predicts the replay oracle's per-event verdict at zero oracle-replay cost.}
\label{tab:related-comparison}
\setlength{\tabcolsep}{4pt}
\renewcommand{\arraystretch}{1.05}
\begin{tabular}{@{}>{\raggedright\arraybackslash}p{0.135\linewidth}>{\raggedright\arraybackslash}p{0.30\linewidth}>{\raggedright\arraybackslash}p{0.13\linewidth}>{\raggedright\arraybackslash}p{0.21\linewidth}>{\raggedright\arraybackslash}p{0.10\linewidth}@{}}
\toprule
Method & Mechanism (cost) & Localizes & Metric (benchmark) & Result \\
\midrule
AGDebugger~\citep{epperson2025agdebugger} & Interactive edit-and-replay (re-exec.\ $+$ human) & message / step & user study ($n{=}14$) & qual. \\
What-if / DoVer~\citep{li2026whatif,ma2025dover} & Replay with interventions (re-exec.) & event / hypothesis & case studies & --- \\
Who\&When judge~\citep{zhang2025whowhen} & LLM-as-judge (no replay) & agent / step & step acc.\ (Who\&When) & $0.14$--$0.52$ \\
AgenTracer-8B~\citep{zhang2025agentracer} & RL tracer; replay for training data & agent / step & step/agent acc.\ (Who\&When) & SOTA ($+18\%$) \\
TRAIL~\citep{deshpande2025trail} & LLM reasoning (no replay) & error step $+$ type & localization (148 tr.) & SOTA $<0.20$ \\
\midrule
\textbf{Ours} & \textbf{Zero-replay oracle prediction (no re-exec., no judge)} & high-effect events & Recall@5 (37 fam., 163k ev.) & \textbf{0.93} \\
\bottomrule
\end{tabular}
\end{table}

\section{Trace Knowledge Representation and Debugging Decision Support}
\label{sec:kr-dss}

\paragraph{From raw logs to an event knowledge graph.}
The first stage compiles a raw multi-agent trace into a structured event knowledge graph $G = (V, E, A)$. Nodes $V$ are trace events (messages, routes, memory writes, tool calls, decision events); edges $E$ encode routing and causal adjacency between events; and the attribute map $A$ assigns each node typed knowledge fields: route-graph position, memory/retrieval persistence, tool-call metadata, uncertainty proxies, and optional latent or hidden-proxy payloads. This representation is the knowledge layer of the system: every downstream component operates on $G$ rather than on raw text, which makes the evidence used for each decision explicit and auditable.

\paragraph{Debugging as budget-bounded decision support.}
Given $G$ and a replay budget $K$, debugging becomes a resource-allocation decision: spend $K$ replay calls on the events most likely to be causally decisive. The predictor is the decision-support model. It maps $G$ to a per-event effect estimate and returns a ranked top-$K$ \emph{replay agenda}---a recommendation of where a human-in-the-loop or automated debugger should invest replay. Calibration metrics (Spearman~$\rho$, AUPRC, reliability bins, ECE; Section~\ref{sec:results-calibration}) provide the confidence estimates a decision-support system needs for budget allocation and triage, so that a low-confidence recommendation can be escalated to active replay rather than trusted blindly.

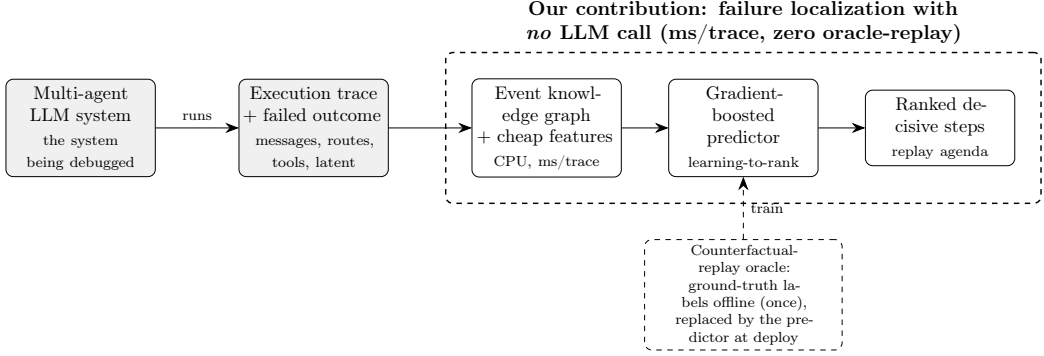
\begin{figure}[t]
\centering
\resizebox{\textwidth}{!}{%
\begin{tikzpicture}[
  font=\footnotesize,
  box/.style={draw, rounded corners, align=center, inner sep=4pt, minimum height=13mm, text width=26mm},
  given/.style={draw, fill=black!6, rounded corners, align=center, inner sep=4pt, minimum height=13mm, text width=26mm},
  ref/.style={draw, dashed, rounded corners, align=center, inner sep=3pt, minimum height=12mm, text width=36mm, font=\scriptsize},
  >={Stealth[length=2.5mm]}
]
\node[given] (sys) {Multi-agent LLM system\\[1pt]{\scriptsize the system being debugged}};
\node[given, right=16mm of sys] (trace) {Execution trace\\ $+$ failed outcome\\[1pt]{\scriptsize messages, routes, tools, latent}};
\node[box, right=16mm of trace] (kg) {Event knowledge graph\\ $+$ cheap features\\[1pt]{\scriptsize CPU, ms/trace}};
\node[box, right=9mm of kg] (gbm) {Gradient-boosted predictor\\[1pt]{\scriptsize learning-to-rank}};
\node[box, right=9mm of gbm] (out) {Ranked decisive steps\\[1pt]{\scriptsize replay agenda}};
\draw[->] (sys)-- node[above,font=\scriptsize]{runs} (trace);
\draw[->] (trace)--(kg);
\draw[->] (kg)--(gbm);
\draw[->] (gbm)--(out);
\node[ref, below=12mm of gbm] (oracle) {Counterfactual-replay oracle:\\ ground-truth labels offline (once),\\ replaced by the predictor at deploy};
\draw[->,dashed] (oracle.north)--(gbm.south) node[midway,right,font=\scriptsize]{train};
\node[draw, dashed, thick, rounded corners, fit=(kg)(gbm)(out), inner sep=5mm] (ourbox) {};
\node[anchor=south, font=\small\bfseries, text width=85mm, align=center] at (ourbox.north) {Our contribution: failure localization with \emph{no} LLM call (ms/trace, zero oracle-replay)};
\end{tikzpicture}%
}
\caption{Where our method sits. We are \emph{given} a recorded execution trace from a multi-agent LLM system and localize the decisive event(s) directly from it: the trace is compiled into an event knowledge graph, cheap CPU features are extracted, and a gradient-boosted learning-to-rank predictor emits a ranked replay agenda---with \emph{no} LLM call. The counterfactual-replay oracle supplies training labels offline (once) and is replaced by the predictor at deployment (zero replay).}
\label{fig:system}
\end{figure}

\paragraph{System architecture.}
Figure~\ref{fig:system} shows where our method sits. The multi-agent LLM system runs the task and leaves an execution trace (often a failure); this trace is our input---we \emph{debug} it, we do not re-solve the task, and we add no LLM call on top. The trace is compiled into the event knowledge graph; cheap features are extracted on CPU in milliseconds; a gradient-boosted predictor scores every candidate event; and a decision-support stage emits the budget-bounded replay agenda. The counterfactual-replay oracle supplies ground-truth labels offline (once) to train the predictor and is replaced by it at deployment---this is what makes the system zero-replay: the expensive oracle is removed from the per-event scoring loop entirely, and no LLM judge is queried, in contrast to LLM-based debuggers.

\section{Method}

The method separates two layers: a deterministic counterfactual-replay \emph{oracle} used only as ground-truth reference, and a \emph{zero-replay predictor} that estimates which events the oracle would mark high-effect without calling it. The predictor is the contribution.

\paragraph{Problem setup and oracle.}
Let a trace be an ordered event sequence $\tau=(e_1,\ldots,e_T)$ with observable fields $o_t$, optional latent fields $z_t$, and outcome $Y(\tau)$. A replay policy $R(\tau,t,i)$ restarts the trace from checkpoint $t$ under intervention $i\in\mathcal{I}_t$ (removing or perturbing a message, route, memory write, tool call, agent, or latent proxy), inducing $P_t^R(y)=\Pr_{i\sim q(\mathcal{I}_t)}[Y(R(\tau,t,i))=y]$. The oracle scores each event by its effect
\[
\Delta(t)=D(P_t^R,\delta_{Y(\tau)})+\lambda_s\,d_s(t)+\lambda_\ell\,d_\ell(t),
\]
an outcome-distribution shift $D$ plus state divergence $d_s$ and trajectory-label divergence $d_\ell$ under deterministic remove-event replay. This is leave-one-out ablation in another vocabulary~\citep{pearl2009causality}, locked across all experiments as reference truth; it is intervention-relative (salient under the chosen $R$, $\mathcal{I}_t$, $q$, $Y$, not an absolute causal time), so we report oracle sensitivity rather than treating one label set as ground truth. Public datasets are converted into replayable traces, not used as task leaderboards~\citep{yang2018hotpotqa,geva2021strategyqa,cobbe2021gsm8k,clark2018arc,narayan2018xsum,austin2021program}.

\paragraph{Zero-replay predictor.}
The predictor never observes $\Delta(t)$ at selection time. Each event is encoded through the replay-artifact fields (text, event type, route edges, memory/retrieval and tool metadata, uncertainty proxies, optional latent fields), the typed attributes of the event knowledge graph $G_\tau$ (Section~\ref{sec:kr-dss}). From these we compute 13 features $\phi$ (graph centrality, routing bottleneck, disagreement, memory/RAG persistence, novelty, uncertainty, latent transitions) with no oracle call and score $S(t)=w^\top\phi(o_{\le t},z_{\le t},G_\tau)-\beta\,C_R(t)$, followed by redundancy-aware top-$K$ selection; with a bounded similarity penalty $F(A)=\sum_{t\in A}S(t)-\gamma\sum_{t<u}\mathrm{sim}(t,u)$ is monotone submodular, so greedy top-$K$ is an appropriate approximation. The fixed-weight and cross-fit \emph{linear} variants give each feature one global weight; since the decisive feature is regime-dependent (Section~\ref{sec:regime}), we also fit a single gradient-boosted tree over the \emph{same} 13 features (depth $3$, $400$ trees, learning rate $0.08$, $\ell_2=1$, minimum leaf $50$) with a \emph{learning-to-rank} objective (target: within-trace rank of $\Delta$) and family-balanced weights. All variants are scored zero-replay and evaluated by five-fold cross-validation grouped by trace and by held-out family, so any gain is attributable to the model over a fixed feature set. $S(t)$ is a pre-replay selector, not a guarantee of latent control or desired-outcome reachability.

\paragraph{Metrics and cost.}
We report rank calibration (Spearman~$\rho$), detection quality (AUPRC and per-trace NDCG against the branchpoint label), and Branch Recall@$K$, with reliability bins and ECE. Selectors sit on a cost-accuracy frontier: \emph{cost} is oracle replay calls per trace (zero for our predictor, the pool size for active replay, $T\cdot|F|$ for the full oracle) and \emph{accuracy} is Branch Recall@$K$. Feature extraction and scoring run in CPU milliseconds against seconds-to-minutes per GPU replay call, so the reported zero cost refers to oracle replay calls, with feature extraction a small bounded CPU cost.

\section{Benchmark Protocol}
For each candidate checkpoint the replay oracle evaluates a bounded intervention family after selection, perturbing event-local information and replaying the downstream trace to estimate outcome, trajectory, behavioral, and state divergence. Table~\ref{tab:eswa-suite} summarizes the benchmark suite.

\scriptsize
\setlength{\tabcolsep}{0pt}
\renewcommand{\arraystretch}{1.08}
\begin{longtable}{@{}>{\raggedright\arraybackslash}p{0.192\linewidth}>{\raggedright\arraybackslash}p{0.192\linewidth}>{\raggedright\arraybackslash}p{0.192\linewidth}>{\raggedright\arraybackslash}p{0.192\linewidth}>{\raggedright\arraybackslash}p{0.192\linewidth}@{}}
\caption{Benchmark suite summary. Public datasets are used as replayable trace sources, not task leaderboard evaluations.}\label{tab:eswa-suite}\\
\toprule
family & trace setting & trace files & task type & replay role \\
\midrule
\endfirsthead
\multicolumn{5}{l}{\scriptsize Continued from previous page.}\\
\toprule
family & trace setting & trace files & task type & replay role \\
\midrule
\endhead
HotpotQA & text/latent/hybrid & 900 & multi-hop QA & public reasoning source for replay traces \\
StrategyQA & text/latent/hybrid & 900 & yes/no reasoning & public reasoning source for replay traces \\
GSM8K & text/latent/hybrid & 900 & arithmetic reasoning & centrality-friendly chain trace source \\
ARC & text/latent/hybrid & 900 & multiple-choice science & route-heavy replay trace source \\
XSum & text/latent/hybrid & 900 & summarization & non-QA hybrid salience source \\
MBPP & text/latent/hybrid tool & 600 & executable tool use & tool/review/verdict replay trace source \\
Hidden/live stress tests & hidden-conditioned/live-agent & 1224 & stress testing & observability and replay-prioritization evidence \\
\bottomrule
\end{longtable}
\normalsize

\section{Results}
\label{sec:results-calibration}

\paragraph{Calibration against the locked oracle.}
Across the 37 trace families (163{,}815 events), the interpretable linear scorer attains Spearman $\rho = 0.58$ ($0.59$ when re-evaluated under the cross-validation harness of Section~\ref{sec:meta}, the figure quoted in the abstract), AUPRC $= 0.61$ (19 points above the $0.418$ positive-class baseline), and ECE $= 0.13$, dominating centrality ($\rho = 0.52$) and routing bottleneck ($\rho = 0.45$); single-feature heuristics (novelty, disagreement, uncertainty, last-$K$) anti-correlate with the oracle ($\rho < 0$), confirming that no single cheap signal suffices. The full per-selector table and reliability diagram are in the supplementary material. This linear scorer is the interpretable baseline; the learning-to-rank predictor of Section~\ref{sec:meta} raises Branch Recall@5 to $0.93$--$0.95$.

\paragraph{Cost-accuracy Pareto frontier.}
Table~\ref{tab:eswa-pareto} reports Branch Recall@5 versus mean oracle replay calls per trace across 37 families. The learned linear scorer at zero replay cost reaches Recall@5 $= 0.73$, recovering $80\%$ of the active-replay oracle's recall ($0.91$ at $11.8$ calls) at zero cost, and dominates the other zero-replay selectors (centrality $0.65$, fixed BranchPoint $0.62$, random $0.45$, last-$K$ $0.29$); the oracle upper bound at cost $5.0$ is $0.89$. At high replay budget active replay strictly dominates, so the contribution is the low-budget regime where exhaustive replay is infeasible.

\begin{table}[t]
\centering
\small
\caption{Cost-accuracy Pareto at $K=5$ (aggregate over 37 families). Cost is mean oracle replay calls per trace; accuracy is Branch Recall@5 against the oracle top-$K$. Zero-replay selectors at cost $0.0$ are directly comparable; active replay and oracle upper bound are reference points.}
\label{tab:eswa-pareto}
\begin{tabular}{lrr}
\toprule
Selector & Cost (calls/trace) & Branch Recall@5 \\
\midrule
\textbf{branchpoint\_learned}   & \textbf{0.0}  & \textbf{0.731} \\
centrality\_only                & 0.0  & 0.646 \\
branchpoint (fixed)             & 0.0  & 0.624 \\
gpu\_embedding\_branchpoint     & 0.0  & 0.599 \\
future\_branchpoint             & 0.0  & 0.543 \\
random\_k                       & 0.0  & 0.449 \\
last\_k                         & 0.0  & 0.294 \\
\midrule
oracle\_upper\_bound (ref.)     & 5.0  & 0.892 \\
active\_replay\_x3 (ref.)       & 11.8 & 0.909 \\
\bottomrule
\end{tabular}
\end{table}

\paragraph{Cross-modality and per-family validation.}
Per-family $K=5$ results across the public latent/hybrid suite (supplementary material) follow the same regime structure: HotpotQA and GSM8K expose centrality-friendly graph regimes, whereas ARC, XSum, and MBPP make tool, latent, uncertainty, or non-QA evidence decisive. The predictor generalizes across modalities while remaining competitive rather than strictly dominant per family. As a secondary efficiency measure, the zero-replay rows obtain about $1.8$--$2.7\times$ visible-token reduction relative to Active Replay $\times 3$ (supplementary).

\section{Learning-to-Rank Meta-Predictor}
\label{sec:meta}

\paragraph{From global weights to regime-conditional interactions.}
The calibration and Pareto results above use the interpretable \emph{linear} scorer, which we retain for per-feature analysis (Sections~\ref{sec:regime},~\ref{sec:ablation}). But a linear scorer assigns each feature one global weight, and Section~\ref{sec:regime} shows the decisive feature is regime-dependent---``use centrality unless the trace is latent- or tool-heavy'' is an interaction, not a weight. Because the operational task is per-trace top-$K$ localization---the quantity the agent failure-attribution literature reports as step-level accuracy or recall~\citep{zhang2025whowhen,deshpande2025trail}---we train a single gradient-boosted tree with a \emph{learning-to-rank} objective (target: each event's within-trace rank of oracle effect) and family-balanced weights. We evaluate under five-fold cross-validation in two schemes: grouping by trace (in-distribution) and, more strictly, by held-out family; because some base traces recur across families, the family split is made trace-disjoint, so it measures genuine unseen-family generalization. We report Branch Recall@5 as the primary metric, NDCG@5 and per-trace~$\rho$ as ranking-quality measures, and global~$\rho$ as a secondary cross-trace calibration figure.

\paragraph{The published linear scorer reproduces, the nonlinear predictor improves.}
Table~\ref{tab:meta-comparison} reports the comparison. Re-evaluated in this harness, the published linear scorer reproduces its reported numbers (Branch Recall@5 $\approx 0.73$, global $\rho \approx 0.59$), which validates the comparison. The learning-to-rank predictor then improves \emph{every} metric under \emph{both} schemes: Branch Recall@5 rises from $0.73$ to $0.95$ in-distribution and $0.93$ on held-out unseen families, NDCG@5 from $0.83$ to $0.94$/$0.92$, per-trace~$\rho$ from $0.47$ to $0.80$/$0.77$, and even global~$\rho$ from $0.59$ to $0.69$/$0.70$ and per-family AUPRC from $0.59$ to $0.77$---all at zero oracle-replay cost. The gains concentrate exactly where linear scoring is blind: on held-out families, MBPP tool traces go from Recall@5 $0.00$ to $1.00$, GSM8K from $0.23$ to $0.98$, XSum from $0.51$ to $0.93$, and ARC from $0.50$ to $0.89$, because the decisive event in those regimes is a tool/latent event with near-zero centrality that a linear weighting cannot surface but tree interactions over the same features can.

\begin{table}[t]
\centering\small
\caption{Learning-to-rank meta-predictor vs.\ the published linear scorer over the \emph{same} 13 features, at zero oracle-replay cost, under five-fold cross-validation grouped by trace (in-distribution) and by held-out family (trace-disjoint). Branch Recall@5 is the primary metric; NDCG@5 and per-trace~$\rho$ measure ranking quality; global~$\rho$ is secondary calibration. The deployed predictor (a single gradient-boosted tree on a within-trace rank target with family-balanced weights) improves every metric under both schemes.}
\label{tab:meta-comparison}
\setlength{\tabcolsep}{6pt}
\resizebox{\textwidth}{!}{%
\begin{tabular}{llcccc}
\toprule
CV grouping & Predictor & Recall@5 & NDCG@5 & per-trace~$\rho$ & global~$\rho$ \\
\midrule
\multirow{3}{*}{by trace (in-distribution)} & Centrality (zero-cost baseline) & 0.736 & 0.796 & 0.465 & 0.518 \\
 & Published linear scorer & 0.733 & 0.831 & 0.478 & 0.595 \\
 & \textbf{Learning-to-rank GBM (ours)} & \textbf{0.945} & \textbf{0.935} & \textbf{0.801} & \textbf{0.690} \\
\midrule
\multirow{3}{*}{by family (purged, unseen)} & Centrality (zero-cost baseline) & 0.736 & 0.796 & 0.465 & 0.518 \\
 & Published linear scorer & 0.730 & 0.828 & 0.472 & 0.586 \\
 & \textbf{Learning-to-rank GBM (ours)} & \textbf{0.926} & \textbf{0.923} & \textbf{0.769} & \textbf{0.701} \\
\bottomrule
\end{tabular}%
}
\end{table}

\paragraph{Trade-offs and objective choice.}
The predictor is near-uniformly better: on held-out unseen families it beats the published scorer on $28$ of $37$ families and loses on only $3$---all high-base-rate StrategyQA variants, and all by a small margin ($0.98 \to 0.94$--$0.98$), so it matches rather than regresses the baseline there. An ablation (supplementary material) shows why the \emph{rank} objective matters: a pointwise variant trained on the raw effect target attains marginally higher cross-trace calibration (global~$\rho$, NDCG@5) but \emph{regresses} on StrategyQA below the published baseline ($0.98 \to 0.69$), whereas the rank objective matches the baseline there while keeping the large gains elsewhere, so we deploy it, since the task is per-trace localization. This also resolves the routing question of Section~\ref{sec:regime}: rather than selecting between centrality and a linear model per trace---which we showed is not cheaply predictable from structural statistics---a single nonlinear predictor \emph{internalizes} the regime-conditional combination and attains the per-regime strengths without any explicit router, at the same zero oracle-replay cost.

\section{Case Study: Model-Authored Live-Agent Traces}
\label{sec:live-agent}

\paragraph{Real agent traces, not adapters.}
To test whether results transfer beyond static dataset adapters, we include six \emph{model-authored} live-agent families across three tasks (ARC, GSM8K, StrategyQA): traces generated by a live LLM agent (Qwen3-1.7B) decoding end to end, up to $144$ traces per task. The full pipeline of Figure~\ref{fig:system} runs unchanged. The learned predictor's rank calibration (Spearman~$\rho$) exceeds centrality on all six families (mean $0.55$ vs.\ $0.29$), and the gap is widest on the most realistic traces (ARC: $0.51$ vs.\ $0.06$): when traces are model-authored rather than adapter-scripted, the structural-only centrality signal degrades while the knowledge-based predictor stays calibrated (full numbers in the supplementary material).

\paragraph{What we do not claim.}
Branch Recall@5 saturates on StrategyQA (every zero-replay selector reaches $1.0$), so $\rho$ and AUPRC are the meaningful metrics there; on the smaller ARC and GSM8K sets, Last-K is competitive on raw Recall@5 though the learned predictor wins on nDCG; live-text task accuracy is unstable across subsets ($0.71$ on ARC, $0.56$ on StrategyQA), consistent with our non-leaderboard framing; and active replay still reaches $1.0$ at $7$--$11$ oracle calls, so the zero-replay predictor's value is in the low-budget regime. Concretely, live generation peaks near $3.3$\,GB of GPU memory and active replay spends $7$--$11$ model executions per trace, whereas the predictor adds a CPU-millisecond pass.

\section{Direct Comparison on Who\&When}
\label{sec:whowhen}

\paragraph{Benchmark and protocol.}
The positioning of Table~\ref{tab:related-comparison} stops short of a shared-leaderboard number because our predictor targets a replay oracle, not human-labeled decisive steps. To place our approach on the field's standard leaderboard, we evaluate on Who\&When~\citep{zhang2025whowhen}, which labels the decisive error step of $184$ failed multi-agent traces ($126$ algorithm-generated, mean $8.7$ steps; $58$ hand-crafted, mean $51.6$ steps). Who\&When's chat-only traces lack our routing, latent, and memory fields, so we cannot transfer the deployed predictor verbatim; instead we apply the \emph{same methodology}---cheap structural and text features (position, length, novelty, agent-disagreement and error-signal proxies) scored by a gradient-boosted ranker, with \emph{no} LLM call at inference---trained on Who\&When's step labels under five-fold cross-validation grouped by trace.

\paragraph{Result.}
Table~\ref{tab:whowhen} reports step-level accuracy (Acc@1), Who\&When's metric. Our LLM-free model, fit in seconds on CPU, reaches Acc@1 $= 0.37$ on the algorithm-generated split and $0.24$ on the hand-crafted split. On the algorithm-generated split this is within $0.01$--$0.02$ of an RL-fine-tuned 8B LLM (AgenTracer-8B, $0.37$~\citep{zhang2025agentracer}) and a frontier LLM judge (Claude-Sonnet-4, $0.39$); on the hand-crafted split it is the highest in the table, above AgenTracer-8B ($0.21$) and Claude-Sonnet-4 ($0.19$), and it beats weaker LLM judges (DeepSeek-R1 $0.30$/$0.07$) and the random and heuristic floors. The relevant comparison is cost: comparable localization with \emph{no} LLM at training or inference, against methods that fine-tune an 8B LLM (GPU-hours) or prompt a frontier model per query. The benchmark's own zero-shot LLM-judge baselines reach only $\le 0.14$ step-level~\citep{zhang2025whowhen}.

\paragraph{What this does and does not show.}
Two caveats qualify the comparison. First, training regimes differ: our model is fit in-domain by 5-fold cross-validation on Who\&When's $184$ traces, whereas AgenTracer is RL-trained on $2000+$ external traces and transferred---so this is a cost--accuracy comparison, not a controlled head-to-head. Second, only the Who\&When-computable subset of our features is available on chat-only traces (no routing, latent, or memory signals), so the number is a lower bound on what the full representation would contribute on richer traces. These results support the paper's central claim: cheap structural signal localizes decisive events at a small fraction of the training and inference cost of LLM-based attribution---a few CPU-seconds of model fitting and no LLM call, versus GPU-hours of fine-tuning or a frontier-model query per step.

\begin{table}[t]
\centering\small
\caption{Step-level localization of the decisive error step on Who\&When (Acc@1, the benchmark's metric; w/o ground-truth-trajectory setting). Our rows (random, heuristic, GBM) are our 5-fold cross-validation on Who\&When; the LLM rows are as reported by AgenTracer~\citep{zhang2025agentracer}. The point is the cost axis: our predictor uses \emph{no} LLM at inference and a few-seconds CPU fit, yet matches an RL-fine-tuned 8B LLM on the algorithm-generated split and exceeds it on the hand-crafted split. Training regimes differ (ours: in-domain CV on $184$ traces; AgenTracer: RL on $2000+$ external traces, transferred), so this is a cost--accuracy comparison, not a controlled head-to-head.}
\label{tab:whowhen}
\setlength{\tabcolsep}{5pt}
\begin{tabular}{lcccc}
\toprule
Method & Inf.\ LLM & Training & Acc@1 (algo.) & Acc@1 (hand) \\
\midrule
Random & --- & --- & 0.135 & 0.052 \\
Longest-step heuristic & $\times$ & --- & 0.182 & 0.190 \\
DeepSeek-R1 (LLM judge) & \checkmark & zero-shot & 0.295 & 0.069 \\
Claude-Sonnet-4 (LLM judge) & \checkmark & zero-shot & 0.388 & 0.190 \\
AgenTracer-8B & \checkmark & RL fine-tune (GPU) & 0.373 & 0.207 \\
\midrule
\textbf{GBM (ours, LLM-free)} & $\times$ & seconds (CPU) & \textbf{0.365} & \textbf{0.241} \\
\bottomrule
\end{tabular}
\end{table}

\section{Ablation and Oracle Robustness}
\label{sec:ablation}
A feature-group ablation and an oracle-sensitivity sweep (supplementary material) keep the predictor auditable: graph evidence is the largest aggregate contributor, latent, routing, and uncertainty evidence remain necessary in non-graph-dominant regimes, and the ranking stays stable under alternative oracle definitions. Desired-outcome branch-discovery results (how many bounded replay interventions rooted at selected checkpoints reach a target outcome) are also in the supplementary; they support branch discovery without claiming guaranteed reachability.

\section{Regime Analysis: When Is the Learned Predictor Worth It?}
\label{sec:regime}

A debugging decision-support system must choose not only \emph{which} events to replay, but \emph{which selector to trust} for a given trace. Across the 37 families, the learned predictor and zero-cost centrality each win a large share (at Branch Recall@5: learned 18, centrality 15, 4 ties), and the split is regime-structured (per-family breakdown in the supplementary). In graph-dominant traces (HotpotQA-style multi-hop, GSM8K-style chains) centrality matches or exceeds the learned predictor (GSM8K: $0.68$ vs.\ $0.51$); in tool/latent-dominant traces (MBPP tool-use, XSum hybrid, ARC latent) the knowledge-based predictor is decisive (MBPP: $1.00$ vs.\ $0.00$; ARC: $1.00$ vs.\ $0.76$). No single selector dominates every family.

\paragraph{A routing test.}
A natural question follows: rather than always running the learned predictor, can one cheaply \emph{route} each trace to centrality when it ``looks graph-dominant,'' capturing the per-regime best for free? We test this directly. We compute ten purely structural statistics from the event knowledge graph (trace length, agent count, route/tool/latent/decision/message fractions, mean route fan-out, reference density, and memory/RAG fraction)---none of which uses any oracle or selector output---and train a logistic regime detector to predict, per family, whether centrality will match or beat the learned predictor. The detector is evaluated leave-one-family-out (LOFO).

The result is negative for deployment (Table~\ref{tab:regime-routing}). The structural statistics carry some signal (the largest absolute point-biserial correlation with the centrality-wins label is $0.34$), and LOFO routing accuracy reaches $0.65$ versus the $0.62$ majority-class rate of always choosing the learned predictor. However, the routed policy (trace-weighted Recall@5 $=0.719$) still underperforms the always-learned default ($0.731$). The error anatomy explains why: a few false-positive overrides on learned-dominant families erase the gains from correctly identifying centrality-friendly regimes. Even a perfect oracle router would gain little: the per-family routing ceiling is only $+0.024$ trace-weighted ($0.755$ vs.\ $0.731$). We therefore do \emph{not} claim an automatic regime router. The regimes are real and interpretable, but they are not reliably separable from structural trace statistics alone at the family grain, so the calibrated learned predictor is the robust default and the apparent ``free lunch'' of falling back to centrality does not materialize. The constructive resolution follows in Section~\ref{sec:meta}: instead of routing \emph{between} a cheap and a learned selector, a single nonlinear predictor over the same features internalizes the regime-conditional combination, recovering the per-regime strengths without any router. The hidden-observability regime is handled conservatively throughout: hidden evidence informs prioritization only, never direct control.

\begin{table}[t]
\centering\small
\caption{Regime-aware routing of the replay selector (Branch Recall@5, aggregate over 37 families). The detector uses only structural trace statistics (no oracle) and is evaluated leave-one-family-out (LOFO); routing accuracy 65\% (majority-class 62\%). Finding: an automatic structural detector does not beat the always-learned default, and the per-family routing ceiling (oracle router) is small; always-learned is the robust default.}
\label{tab:regime-routing}
\begin{tabular}{lrr}
\toprule
Replay-selector policy & Recall@5 (unweighted) & Recall@5 (trace-weighted) \\
\midrule
Always centrality (zero-cost baseline) & 0.699 & 0.646 \\
Always learned (main predictor) & 0.775 & 0.731 \\
\textbf{Regime-routed (LOFO, ours)} & \textbf{0.762} & \textbf{0.719} \\
Oracle router (ceiling, ref.) & 0.811 & 0.755 \\
\bottomrule
\end{tabular}
\end{table}

\section{Limitations and Threats to Validity}

\paragraph{Oracle definition is intervention-relative.}
The locked oracle uses remove-event interventions only. Branchpoint labels would change under correction, rerouting, swap, or flip interventions. The zero-replay prediction framework remains valid under a different oracle but would require recalibration. A complementary line of work on stochastic, multi-intervention branch discovery (with Future Plasticity Index oracles) is left for future work.

\paragraph{Deterministic replay.}
Replay functions are deterministic. Real LLM-agent traces can be stochastic. Our predictor is well-defined under deterministic replay; stochastic extension requires confidence-interval reporting on oracle effects and is out of scope.

\paragraph{Synthetic and adapter mix.}
Public-derived traces (HotpotQA, StrategyQA, GSM8K, ARC, XSum, MBPP) are adapters from QA, reasoning, summarization, and code datasets, not native multi-agent benchmarks. Calibration may not transfer perfectly to production agent systems. We report cross-family generalization as evidence of robustness, not as guarantee of production fit.

\paragraph{Centrality is competitive in graph-dominant regimes.}
The regime analysis identifies graph-dominant traces where simple centrality matches the learned predictor at zero cost. Our contribution is not universal selector dominance but a calibrated predictor that is the robust default across the suite on aggregate calibration and Pareto position; we further show (Section~\ref{sec:regime}) that cheaply routing to centrality from structural trace statistics does not beat this default, so the learned predictor should be preferred.

\paragraph{Hidden-state controllability is not claimed.}
Hidden-prefix probes report sensitivity, not control. We report 1 strong-positive, 2 weak-positive, 6 sensitivity-only verdicts across 9 settings. This is an observability-gap result, not a control claim.

\paragraph{Active replay dominates at high budget.}
At sufficiently high replay budget, active replay strictly dominates the zero-replay predictor in accuracy. Our contribution is the low-budget regime, where paying $O(T \cdot |F|)$ oracle calls is infeasible and approximate ranking is the only practical option.

\section{Reproducibility, Data Availability, and AI Use}
The submission package includes one-command regeneration of the main tables, a manifest linking experiments to artifact CSVs, a claim-verification script, bootstrap confidence-interval outputs, and a conservative claim audit. The data availability statement, generative AI declaration, and declaration of interest are supplied as separate submission files. Generative AI assistance was used for manuscript preparation and code-editing support, but not as a source of experimental results; numerical claims are regenerated from local artifacts and checked by scripts.

\section{Conclusion}
We reframe counterfactual replay in multi-agent LLM traces as a knowledge-based decision-support problem rather than an oracle-design problem. Traces are compiled into an event knowledge graph; given a fixed oracle as ground-truth reference, BranchPoint-Latent predicts which events the oracle would mark high-effect without paying the replay cost, and a decision-support stage turns those predictions into a budget-bounded replay agenda. Across 37 trace families (163{,}815 events), an interpretable linear scorer reaches Branch Recall@5 $= 0.73$; a single learning-to-rank gradient-boosted predictor over the same features, capturing regime-conditional interactions, raises this to $0.93$ on held-out unseen families and $0.95$ in-distribution (NDCG@5 $0.83\!\to\!0.92$) at zero oracle-replay cost, beating the published scorer on every metric. The contribution is not a new oracle or a universally dominant selector, but a calibrated, auditable decision-support system grounded in an explicit trace knowledge representation, positioned on the cost-accuracy frontier, with reproducible artifacts and conservative claim boundaries. Extending the framework to stochastic replay and richer intervention families is the natural next step.

\bibliographystyle{elsarticle-num}
\bibliography{branchpoint_latent_refs}

\end{document}